\documentstyle[11pt,moriond,epsfig]{article}

\def\be{\begin{equation}}
\def\ee{\end{equation}}
\def\bea{\begin{eqnarray}}
\def\eea{\end{eqnarray}}

\newcommand {\ipb}              {\mbox{pb$^{-1}$}}

\newcommand {\sm}               {Standard Model}

\newcommand {\mee}               {{\mathrm e}^+ {\mathrm e}^-}

\newcommand {\sele}             {\tilde{\mathrm e}}
\newcommand {\sell}             {\tilde{\ell}}

\newcommand {\stau}             {\tilde{\tau}}

\newcommand {\nt}               {\tilde{\chi}^0}

\newcommand {\stopm}         {\tilde{\mathrm{t}}_{1}}

\newcommand {\stopl}         {\tilde{\mathrm{t}}_{\mathrm L}}
\newcommand {\stopr}         {\tilde{\mathrm{t}}_{\mathrm R}}

\newcommand {\neutralino}    {\tilde{\chi }^{0}_{1}}
\newcommand {\neutrala}      {\tilde{\chi }^{0}_{2}}

\newcommand {\chargino}      {\tilde{\chi }^{\pm}_{1}}
\newcommand {\charginos}     {\tilde{\chi }^{\pm}_{1,2}}
\newcommand {\charginop}     {\tilde{\chi }^{+}_{1}}

\newcommand {\charginom}     {\tilde{\chi }^{-}_{1}}

\newcommand {\Zboson}        {{\mathrm Z}^{0}}

\newcommand {\mstop}         {m_{\stopm}}

\newcommand {\mchi}          {m_{\neutralino}}
\newcommand {\mchip}         {m_{\chargino}}

\newcommand {\Rparity}          {$R$-parity}
\newcommand {\sbotx}            {\tilde{\mathrm{b}}}
\newcommand {\sbotm}         {\tilde{\mathrm{b}}_{1}}
\newcommand {\sbots}         {\tilde{\mathrm{b}}_{2}}

\newcommand{\lb}             {$\lambda$}

\begin{document}


\vspace*{4cm}
\title{SUSY SEARCHES AT LEP}

\author{ S. BRAIBANT
}

\address{CERN, EP Division, CH-1211 Geneva 23, Switzerland}

\maketitle\abstracts{
Between 1995--2000, the LEP $\mee$ collider has been operated 
above the Z$^0$ peak, at centre-of-mass energies $\sqrt{s} =$ 130--209~GeV. 
Searches for supersymmetric particles 
have been performed using these data samples. The results from
the four LEP experiments have been combined.   
Model independent limits on the pair-production cross-sections 
of supersymmetric particles and
constraints on their masses are presented in the context 
of the Minimal Supersymmetric 
Standard Model (MSSM) and in the context of gauge-mediated
supersymmetry breaking models (GMSB).
Results assuming an \Rparity\ violating scenario 
are also reviewed. 
}

\section{Introduction}

Between 1995--2000, the LEP $\mee$ collider has been operated 
at centre-of-mass energies $\sqrt{s}=$ 130--209~GeV. 
The total integrated luminosity collected 
at these energies 
by each of the four LEP experiments, 
ALEPH, DELPHI, L3 and OPAL, is about 700~$\ipb$.

All results described here constitute a representative selection 
of the combination of results from the four
LEP experiments.
These ``LEP combinations'' are preliminary and  
a full description can be found in~\cite{ref:ALL}.
In most of the searches, some candidates are selected
by the analyses but their number is compatible with the expected
background from \sm\ (SM) processes.
Since no significant excesses with respect to the \sm\ background 
were observed, 95\% confidence level (C.L.) upper limits 
on the production cross-section were computed and mass limits were 
derived.

\section{Searches for SUSY Particles in the MSSM framework}

In SUSY models, 
each particle is accompanied by a
supersymmetric partner whose spin differs by half a unit.
It is often assumed that
\Rparity\ (R$\equiv$(-1)$^{{\mathrm 2S+3B+L}}$) is conserved and
that the lightest neutralino, $\nt_1$, is the lightest
supersymmetric particle (LSP). \Rparity\ conservation implies that
SUSY particles are always pair-produced and always decay, through
cascade decays, to ordinary particles and $\neutralino$. Moreover
the $\neutralino$ is stable and
escapes detection due to its weakly interacting nature.
A characteristic signature of all events containing SUSY
particles is therefore missing energy and momentum.


Each lepton has two scalar partners, the right and left-handed 
charged scalar leptons (sleptons),
denoted $\sell_{\mathrm R}$ and $\sell_{\mathrm L}$, according 
to their helicity states. 
The dominant slepton decay mode is~: $\sell^\pm \rightarrow \ell^\pm + \nt_1$.
The event topologies are two acoplanar leptons and missing energy.

\begin{figure}[t] 
\centering
\begin{tabular}{cc}
\epsfig{file=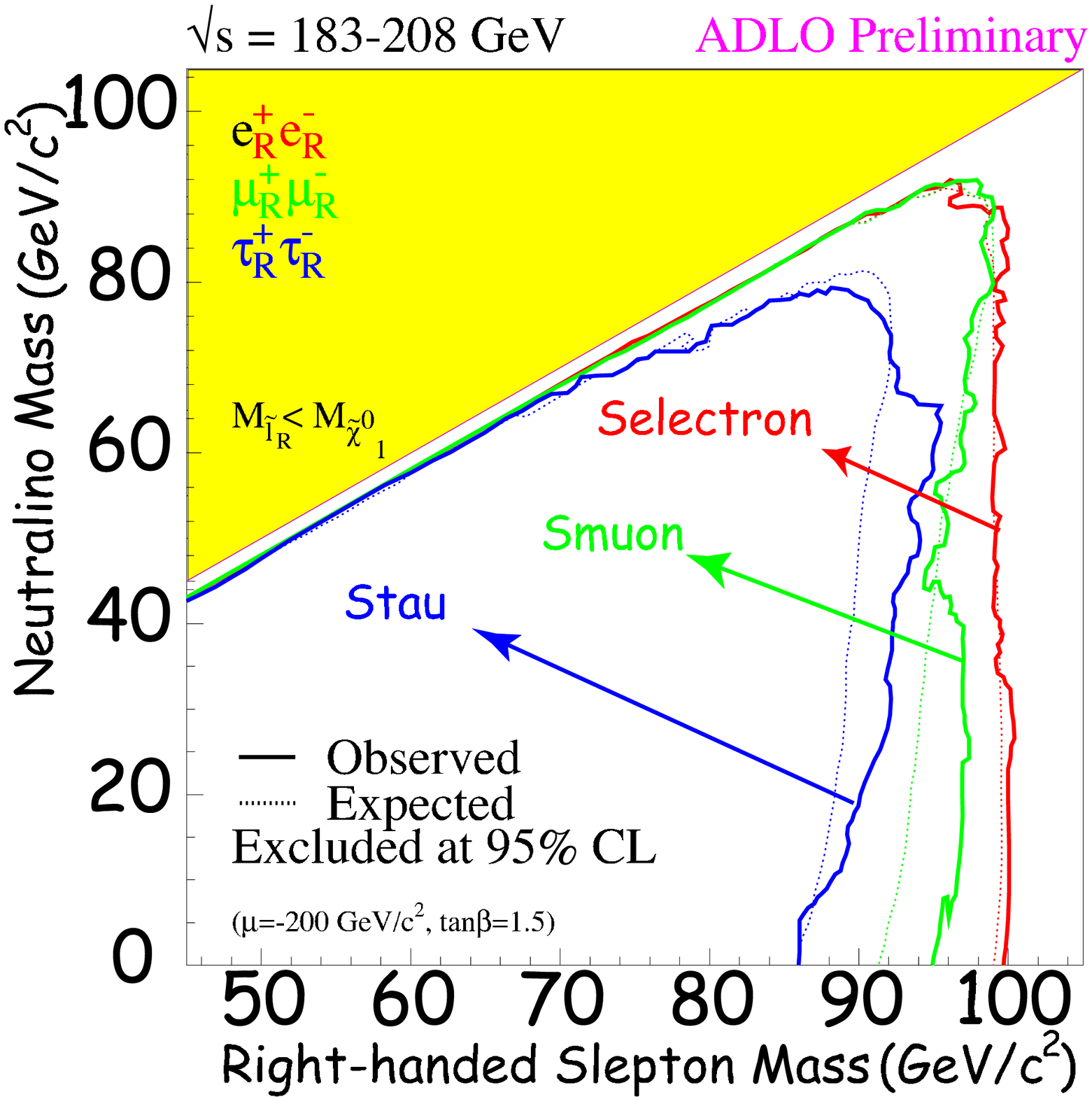,height=6cm} 
 & \epsfig{file=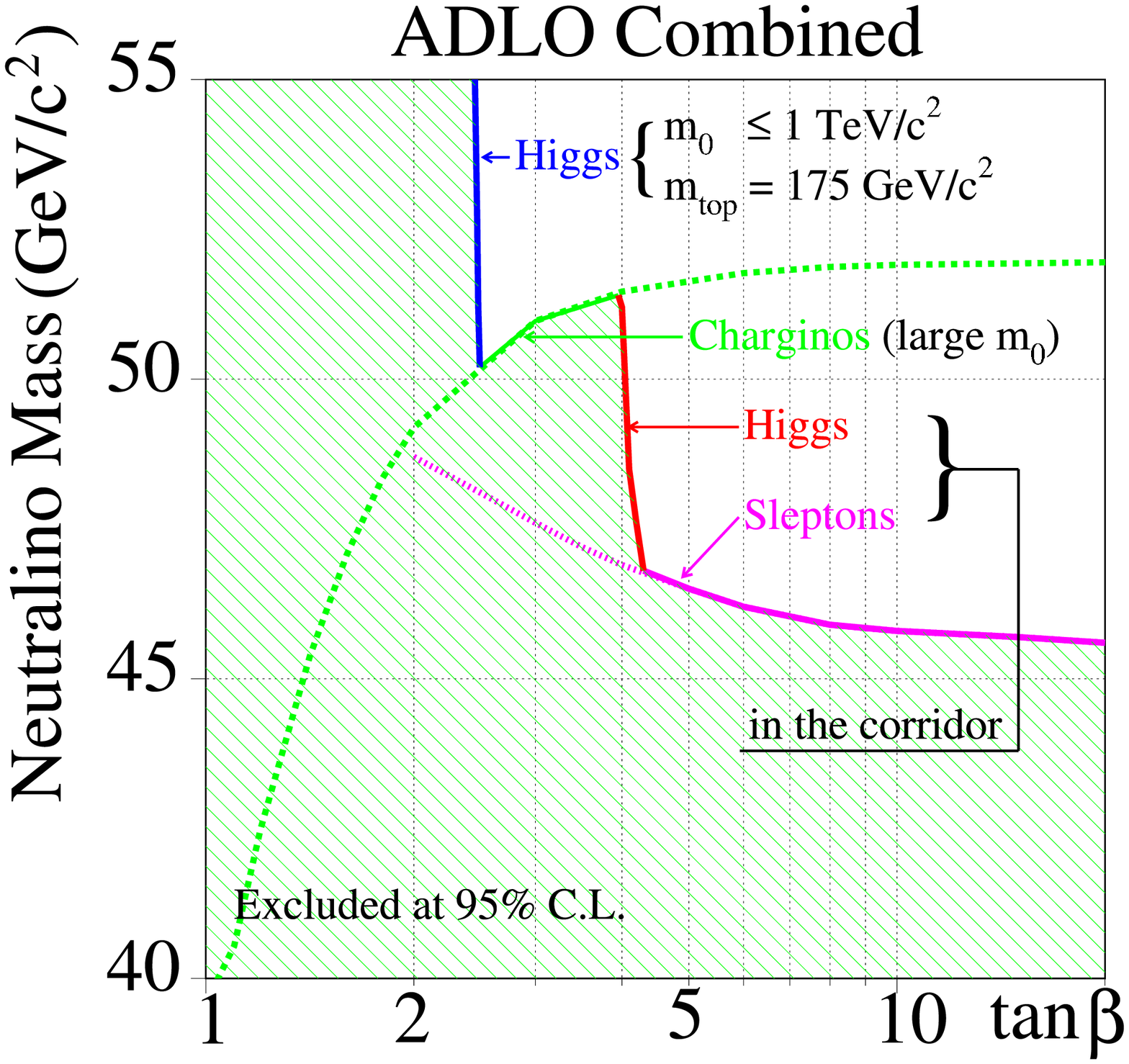,height=6.3cm} \\ 
\end{tabular}
\vspace{10pt}
\caption{
Left plot: 95\% C.L. exclusion regions (solid lines) 
for pair-produced right-handed selectrons, smuons and staus 
as a function of the $\neutralino$ mass. For the 
selectron case,
tan $\beta =$ 1.5 and $\mu = - 200$~GeV are used. 
The dotted line represents 
the expected limits.
Right plot: 95\% C.L. $\neutralino$ mass lower limit as a function 
of $\tan \beta$. }
\label{fig:sleptonfig}
\end{figure}

Exclusion regions
for pair-produced right-handed selectrons, smuons and staus 
are shown in 
Fig.~\ref{fig:sleptonfig} (left plot). 
A right-handed selectron with a mass smaller than 99.6~GeV, 
a right-handed smuon with a mass smaller than 94.6~GeV 
and right-handed stau with a mass smaller than 85.9~GeV 
are excluded.

Searches for scalar top and bottom quarks (stop and sbottom) 
were also performed. Because of the large mass splitting by left-right mixing, 
$\stopm = \stopl \cos \theta_{mix} + \stopr \sin \theta_{mix}$, the 
lowest mass eigenstate, $\stopm$, could be the lightest 
charged supersymmetric particle.
The stop quark pair-production cross-section depends on the 
stop mass, $\mstop$, and the mixing angle $\theta_{mix}$.
If the stop quark is assumed to be lighter 
then every other charged sparticle, the dominant 
decay mode is the 2-body flavour changing  
decay $\stopm \rightarrow {\mathrm c} + \neutralino$.
The event topologies would therefore be two acoplanar jets with
missing energy.
For $\theta_{mix}=56^o$ and assuming the 2-body decay, 
a stop with a mass smaller than 95~GeV is excluded.
For large $\tan \beta$, there could be a large mixing also between the
right- and left-handed $\sbotx$ quarks, 
resulting in two states, $\sbotm$ and $\sbots$. The lowest lying state,
$\sbotm$ would decay primarily to $\sbotm \to {\mathrm b} + \neutralino$
leading to a topology identical to the pair-production of
stop quarks followed by 2-body decay. 
For $\theta_{mix}=68^o$, 
a sbottom with a mass smaller than 94~GeV is excluded.


The charginos, $\charginos$, could be pair-produced at LEP either through 
$\gamma$ or $\Zboson$ exchange in the $s$-channel or through 
sneutrino exchange in the
$t$-channel. The production cross-section could be fairly large 
and therefore the search for charginos was one of the most appealing 
SUSY searches. 
The $\chargino$ could decay into a $\neutralino$ and an ordinary lepton~:
$\chargino \rightarrow \neutralino l^\pm \nu$ (leptonic decay), or into a
neutralino and a quark pair~: $\chargino \rightarrow \neutralino 
{\mathrm q\overline{q\prime}}$ (hadronic decay) through a virtual W*, 
slepton or scalar quark emission.
The experimental signature for $\chargino$ production is therefore~:
a) two acoplanar leptons, b) one lepton plus jets or c) multi-jets; all
these topologies share the characteristic of a large missing energy carried
away by the neutralinos. The most challenging search is when the mass
difference between the $\chargino$ and $\neutralino$ 
($\Delta m = \mchip -\mchi$) is small, because the 
event visible energy also becomes very small.
Detailed searches for charginos 
in the small $\Delta m$ region have been performed. In the case of 
a very small mass 
difference ($\Delta m \leq 0.1$~GeV), the charginos would be  
quasi-stable charged particles. 
The searches are based on the specific
ionisation loss measurement, dE/dx, provided by the central 
detectors of each experiment. 
No evidence for chargino production has been observed by the four 
LEP collaborations in any of the $\Delta m$ regions. 
The 95\% C.L. lower limits on the $\chargino$ mass 
obtained for the case of a heavy sneutrino and 
$\Delta m \geq$ 10~GeV are close to the kinematic limit.

The neutralinos could be pair-produced through $s$-channel 
virtual $\Zboson$ exchange 
or $t$-channel scalar electron exchange. 
Alternatively one can look directly for the production of a 
$\neutrala \neutralino$ pair. The $\neutrala$ could then decay into
$\neutralino l^+l^-$, $\neutralino \nu\overline{\nu}$ or 
$\neutralino {\mathrm q\overline{q}}$, through a virtual $\Zboson$, 
Higgs boson, slepton or squark exchange.
The event topologies are similar to those studied for 
$\charginop \charginom$ events.

From these studies, a 95\% C.L. lower mass limit of $m_{\neutralino} >$46 GeV 
was derived for any value of $\tan \beta$ (right plot of 
Fig.~\ref{fig:sleptonfig}). Since the LSP could be a good candidate for 
dark matter, this lower mass bound is interesting also 
from a cosmological point of view.

\section{Searches for Gauge Mediated Supersymmetry Breaking (GMSB) Signatures}

\begin{figure}[t] 
\centering
\begin{tabular}{cc}
\epsfig{file=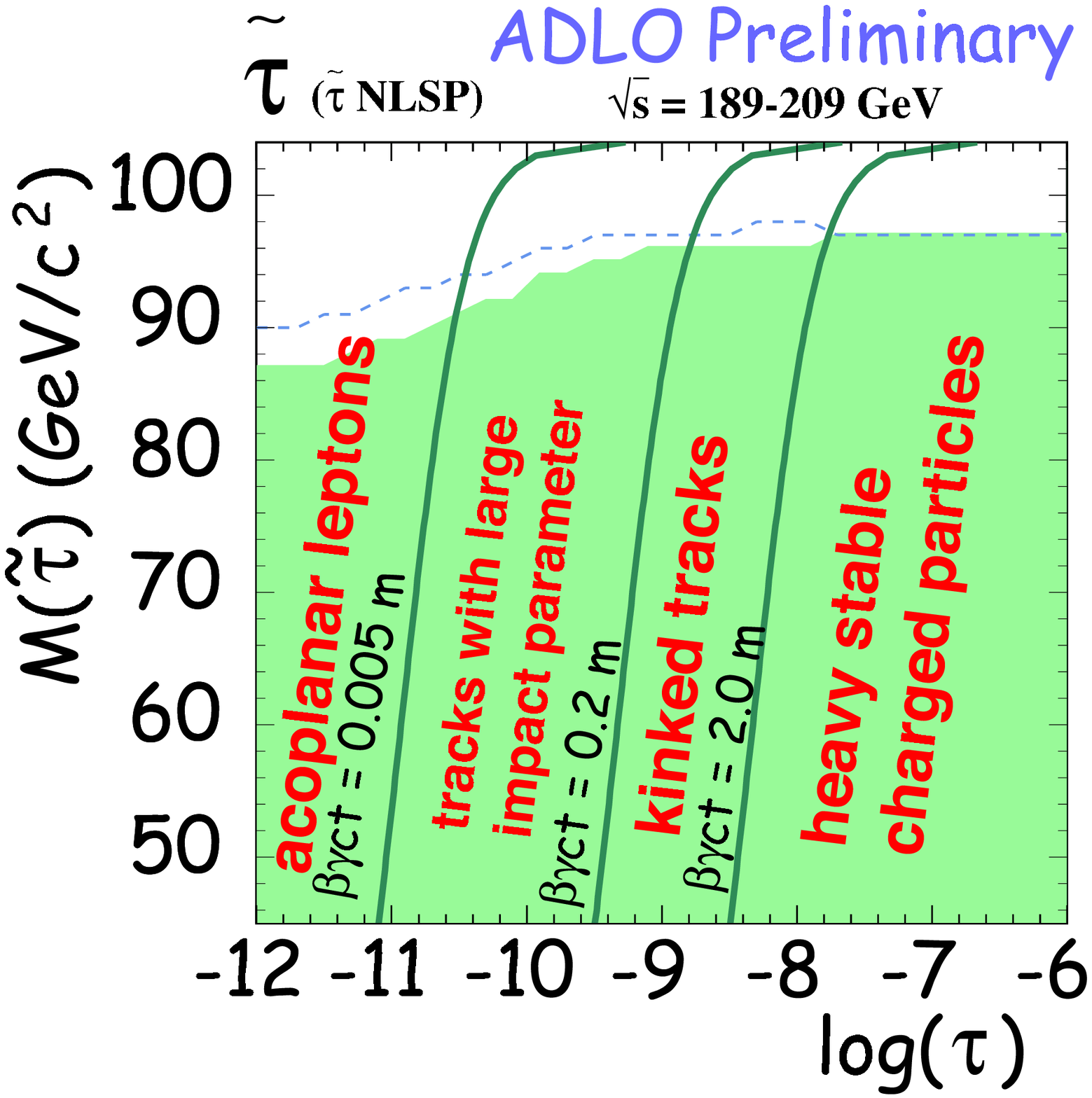,height=6cm} 
& \epsfig{file=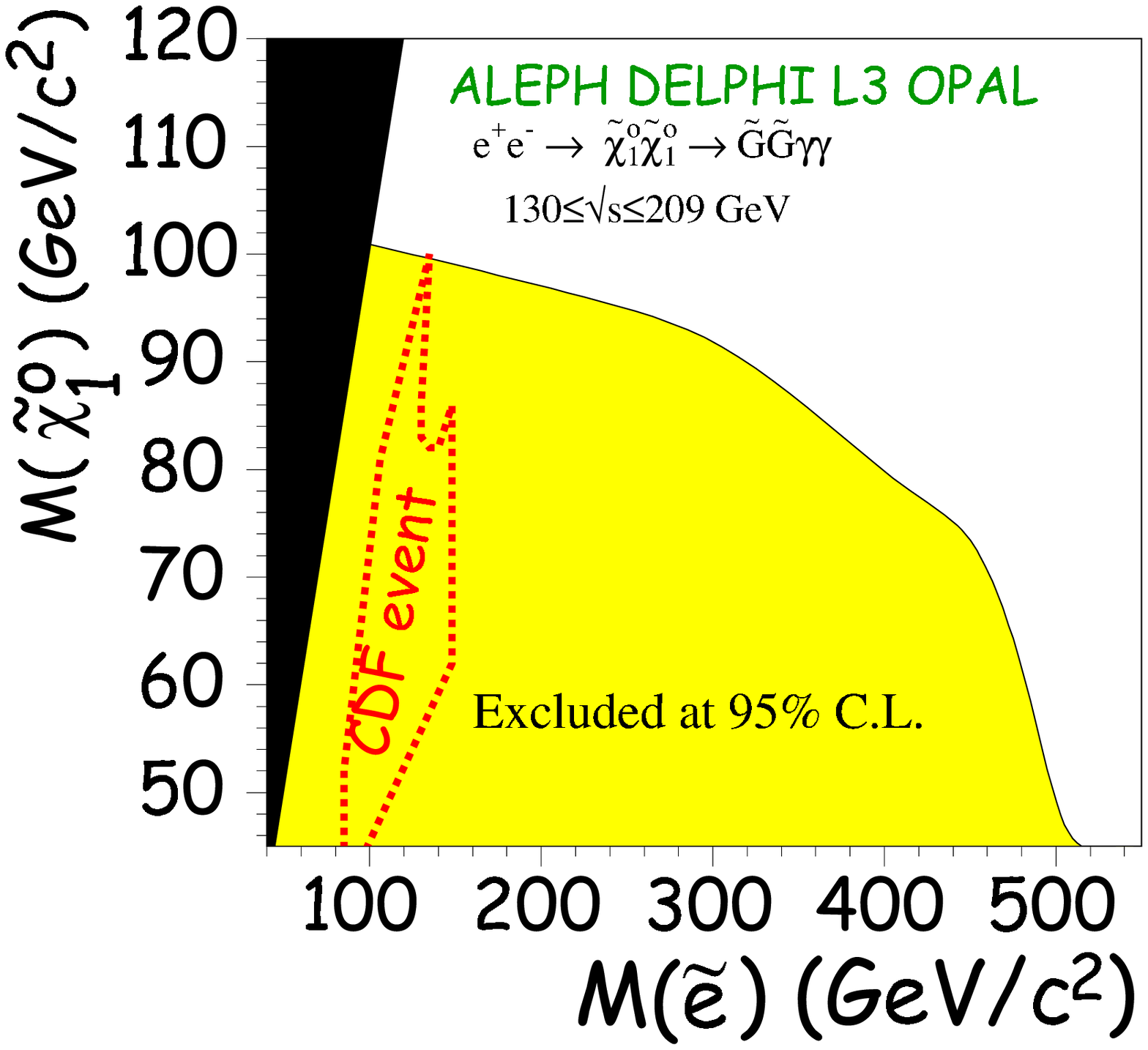,height=6cm} \\ 
\end{tabular}
\vspace{10pt}
\caption{
Left plot: 95\% C.L. excluded mass regions for right-handed staus (NLSP) as a 
function of the NLSP lifetime combining four
different searches.  
Right plot: 95\% C.L. excluded region in 
the ($m_{\sele_R}$, $m_{\neutralino}$) 
mass plane assuming 
a $\neutralino$ NLSP; this limit is 
derived from the search for acoplanar photons assuming a zero lifetime NLSP.}
\label{fig:gmsb}
\end{figure}

In the GMSB framework, supersymmetry is broken via the usual 
gauge interactions in a hidden sector, which 
couples to the visible sector of the \sm\ and SUSY 
particles via a messenger sector. 
The supersymmetric partner of the graviton, the gravitino 
$\tilde{G}$ is assumed to be the LSP, and the next-to-LSP (NLSP) could either 
be the lightest neutralino, $\tilde{\chi}_1^0$, 
or a right-handed slepton, $\tilde{l}_R$. 
The NLSP decay length is unconstrained and 
all possible decay lengths between zero and infinity had to be considered,
suggesting to explore many different final state topologies.
Fig.~\ref{fig:gmsb} (left plot)  shows the 95\% C.L.
excluded $\stau_R$ mass as a function of the lifetime, assuming a 
$\stau_R$ NLSP and combining searches exploring various regions of the NLSP 
lifetime. 
The very short lifetime range is 
covered by the search for events with a pair of acoplanar leptons; 
the intermediate lifetime
range is covered by the searches for events containing tracks 
with large impact parameters and kinks; 
and the long lifetime range is covered by the search for heavy 
stable charged particles. 
Fig.~\ref{fig:gmsb} (right plot) shows the 
95\% C.L. excluded region in the ($\sele_R$, $\neutralino$) 
mass plane assuming 
a $\neutralino$ NLSP; this limit is 
derived from the search for events with acoplanar 
photons and assuming a zero lifetime NLSP.

\section{Searches for \Rparity\ Violation Decays of Supersymmetric Particles}

If \Rparity\ is violated, the sparticles could decay directly to \sm\
particles and any sparticle could be the LSP.
The topologies differ significantly from the
ones with conserved \Rparity.

With the MSSM particle content,  \Rparity\ violating interactions are 
parametrised with a gauge-invariant super-potential that includes the 
following Yukawa coupling terms: 
\bea
W_{RPV}  = 
    \lambda_{ijk}      L_i L_j {\overline E}_k
 +  \lambda^{'}_{ijk}  L_i Q_j {\overline D}_k
 +  \lambda^{''}_{ijk} {\overline U}_i {\overline D}_j {\overline D}_k, 
\label{lagrangian}
\eea
where $i,j,k$ are the generation indices of the super-fields 
$L, Q,E,D$ and $U$. $L$ and $Q$ are  
respectively the lepton and quark left-handed doublets.
$\overline E$, $\overline D$ and $\overline U$ are respectively 
the right-handed 
singlet charge-conjugate super-fields for the charged 
leptons and down- and up-type quarks.
This makes a total of 45 parameters in addition to those 
of the \Rparity\ conserving MSSM.


\begin{figure}[t] 
\centering
\begin{tabular}{cc}
\epsfig{file=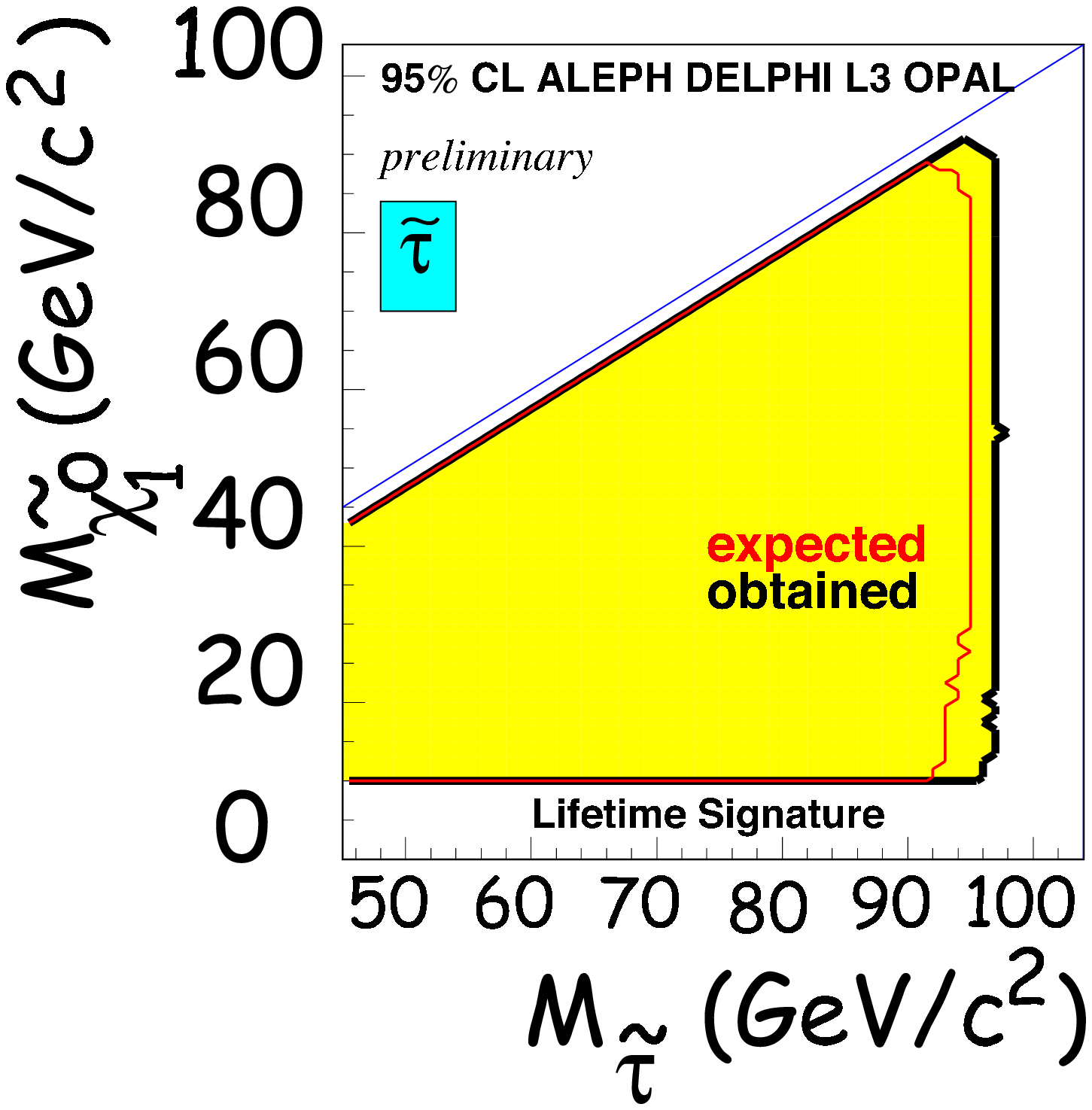,height=6.0cm} 
& \epsfig{file=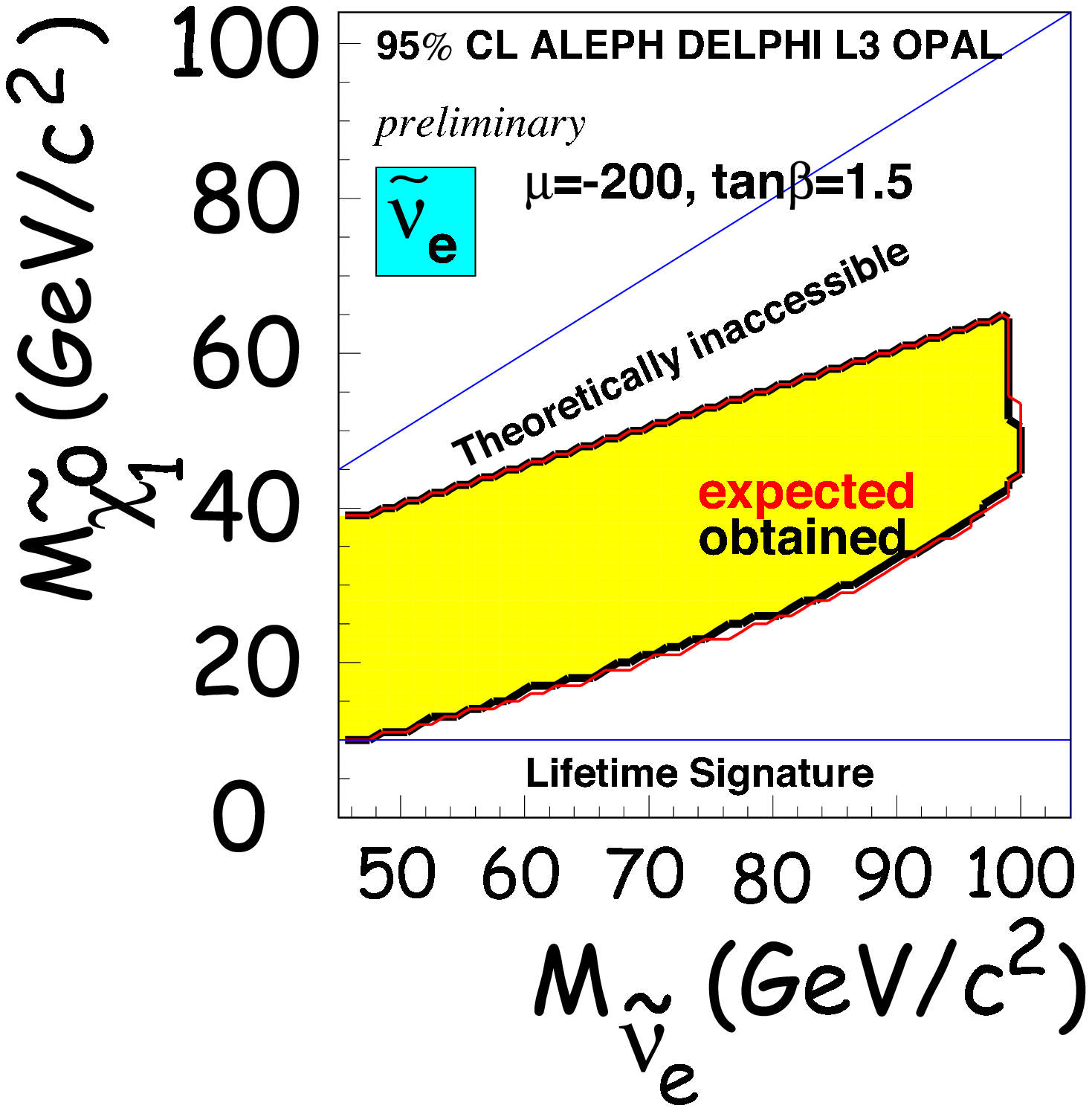,height=6.0cm} \\ 
\end{tabular}
\vspace{10pt}
\caption{
95\% C.L. lower mass limits for a right-handed selectron (left plot)
and  for a sneutrino (right plot).
The exclusions are shown for a \lb\ coupling and for 
tan $\beta =$ 1.5 and $\mu = - 200$~GeV.
}
\label{fig:rpvfig}
\end{figure}

The left plot of Fig.~\ref{fig:rpvfig} shows the 95\% C.L. 
lower mass limits for right-handed pair-produced staus  
while the right plot shows the 95\% C.L. limits for pair-produced 
sneutrinos.  
The exclusions are shown for a \lb\ coupling and for 
tan $\beta =$ 1.5 and $\mu = - 200$~GeV.

\section{Conclusions}
LEP has been a great success until its very end allowing 
a multitude of searches for new particles. These searches have  
been performed using a total integrated luminosity of about 700~$\ipb$ 
per experiment, at centre-of-mass energies up to 209~GeV. 
A number of preliminary combinations of these searches for new particles have 
been performed and interpreted in various models. 
No significant evidence for new physics is observed. Very stringent 
95 \% C.L. limits on cross-sections and sparticle masses have been computed.

\end{document}